# Single-Ferroelectric Memcapacitor-Based Time-Domain Content-Addressable Memory for Highly Precise Distance Function Computation

Minjeong Ryu, Jae Seung Woo, Yeonwoo Kim, and Woo Young Choi

*Abstract*— Single ferroelectric memcapacitor-based time-domain (TD) content-addressable memory (CAM) is proposed and experimentally demonstrated for high reliability and density. The proposed TD CAM features the symmetric capacitance-voltage characteristics of a ferroelectric memcapacitor with a gated p-i-n diode structure. This CAM performs search operations based on the variable capacitance of cells. The propagation delay of the TD CAM output signal is linearly correlated with the Hamming distance (HD) between input and output vectors. The proposed TD CAM array exhibits exceptional reliability in HD computation and in-memory search tasks owing to this linearity, considerably outperforming the conventional nonlinear voltage-domain CAM.

*Index Terms*—ferroelectric memcapacitor, time-domain (TD) content-addressable memory (CAM), capacitive in-memory computing.

## I. INTRODUCTION

CONTENT-ADDRESSABLE-MEMORY-BASED (CAM-based) in-memory nearest neighbor (NN) search is advancing applications in edge artificial intelligence (AI), including memory-augmented neural networks for one- or few-shot learning and hyperdimensional computing [1]–[5]. The key function of CAM in this type of associative search is the parallel computation of the Hamming distance (HD), which measures the degree of mismatch between a query vector and stored vectors. However, conventional voltage-domain (VD) CAMs suffer from a nonlinear relationship between the HD and voltage output, as the search operation relies on Ohm's law and Kirchhoff's laws [1]–[6]. Resistive HD computing schemes encounter comprehensive constraints in system-level performance, such as degraded accuracy, lower sensing margin, and inevitable area/power overhead of peripheral circuits arising from nonlinearity [6]–[10].

To overcome these problems, a time-domain (TD) CAM has been proposed, encoding bits based on cell capacitance rather than conductance [6]–[10]. By adjusting the variable capacitance instead of the driving current, the HD computation results can be linearly converted into the time delay of the output signal [6]–[8], [11]. In addition to the high precision of NN searches associated with linearity, capacitive computing is advantageous because of its remarkable energy efficiency due to its negligible static current and high resilience to device variations [6], [13]–[20]. Moreover, TD CAM has compact peripheral circuitry owing to its digital compatibility [6]–[12].

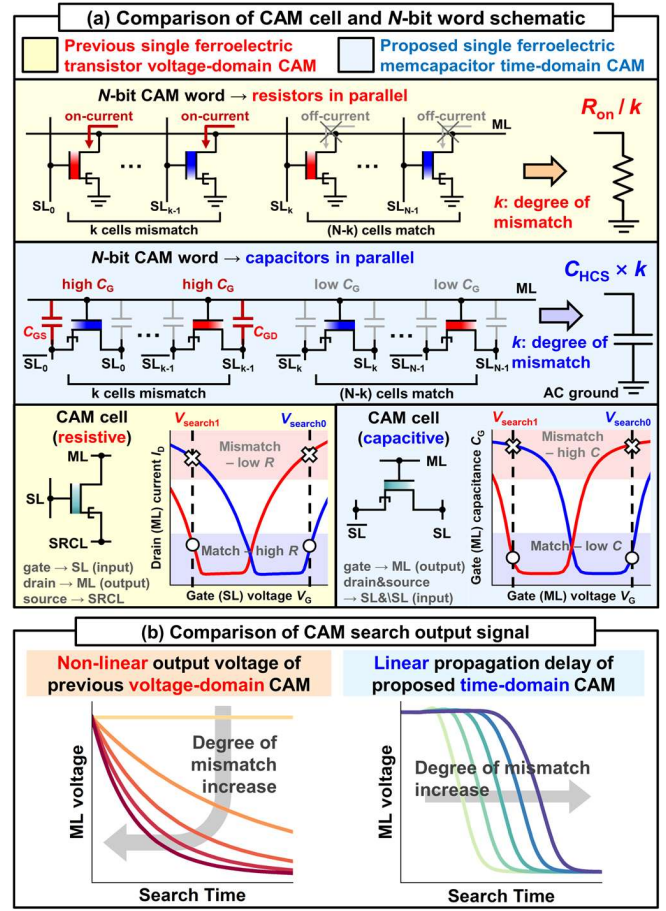

Fig. 1. Comparison of the previous one-transistor VD CAM with the proposed one-memcapacitor TD CAM (a) in terms of operation mechanism and (b) search output signal.

This study proposes a novel TD CAM based on a single ferroelectric memcapacitor with symmetric capacitance–

This work was supported in part by Samsung Research Funding & Incubation Center of Samsung Electronics under Project Number SRFC-TA2103-01 and in part by the NRF of Korea funded by the MSIT under Grant RS-2024-00402495, NRF-2022M3I7A1078544 (PIM Semiconductor Technology Development Program). (*Corresponding Author: Woo Young Choi.*)

M. Ryu, J. S. Woo, Y. Kim, and W. Y. Choi are with the Department of Electrical and Computer Engineering and the Inter-university Semiconductor Research Center (ISRC), Seoul National University, Seoul 08826, Republic of Korea (e-mail: wooyoung@snu.ac.kr).

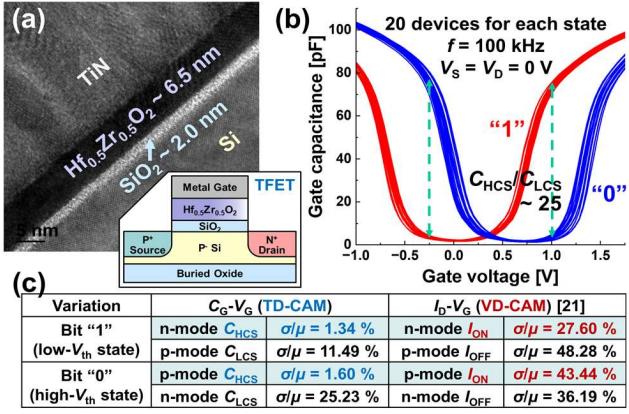

Fig. 2. (a) Cross-sectional TEM image of the gate stack of the fabricated ferroelectric memcapacitor and its schematic with the TFET structure. (b) Measured *C-V* curves (gate area = 10000 μm$^2$) and (c) extracted coefficient of variation from device characteristics. Two shaded rows $C_{HCS}$ and $I_{ON}$ plays the main role in driving the CAM response.

TABLE I
WRITE AND SEARCH BIAS SCHEMES FOR THE PROPOSED TD CAM

| Operation | Write | | Search | |
|---|---|---|---|---|
| Bit | "1" | "0" | "1" | "0" |
| [a]Node and pulse value | $V_{ML}$ = 6.5V, 500 ns | $V_{ML}$ = -6.5 V, 500 ns | $V_{SL} = V_{\backslash SL}$ = 0.3 V | $V_{SL} = V_{\backslash SL}$ = -1.0 V |

[a]Nodes not indicated are all grounded during specified operation.

voltage (*C-V*) characteristics, as illustrated in Fig. 1a. In this configuration, the high- ($C_{HCS}$) and low-capacitance states ($C_{LCS}$) of the ferroelectric memcapacitor indicate the mismatch and match states, respectively. This is analogous to previous VD CAM designs based on single ferroelectric transistors with ambipolar current–voltage (*I-V*) characteristics, where the on-current ($I_{ON}$) and off-current ($I_{OFF}$) of the cell correspond to the mismatch and match states, respectively [21], [22]. The proposed scheme addresses the nonlinearity problem of the previous single-transistor VD CAM, as depicted in Fig. 1b. The propagation delay of the proposed TD CAM output is modulated linearly with the HDs as the computed results are transformed into the summation of the capacitors in parallel.

## II. RESULTS AND DISCUSSION

### A. Ferroelectric Memcapacitor Cell Characteristics

The ferroelectric memcapacitor in the TD CAM cell is based on the structure of p-i-n tunnel field-effect transistors (TFETs), as shown in the inset of Fig. 2a. The TD CAM cell configuration is illustrated in Fig. 1a, where the gate, drain, and source of the device are connected to the match line (ML), search line (SL), and search line bar (\SL), respectively. The cross-sectional view in Fig. 2a illustrates the metal-ferroelectric-insulator-semiconductor (MFIS) gate stack of a device fabricated using a complementary metal oxide semiconductor (CMOS) gate-first process [21], [23]. Fig. 2b presents the measured symmetric *C-V* characteristics, wherein a high gate-to-source capacitance $C_{GS}$ (p-type inversion), low gate capacitance (depletion), and high gate-to-drain capacitance $C_{GD}$ (n-type inversion) appear in sequence [14], [24], [25]. This completely differs from the asymmetric nature of MOSFET *C-V* curves [16]–[20]. Additionally, as the $C_{LCS}$ is determined by silicon depletion, our proposed memcapacitor offers a greater dynamic range and sensing margin compared to metal-ferroelectric-metal-based (MFM-based) schemes [15]–[18].

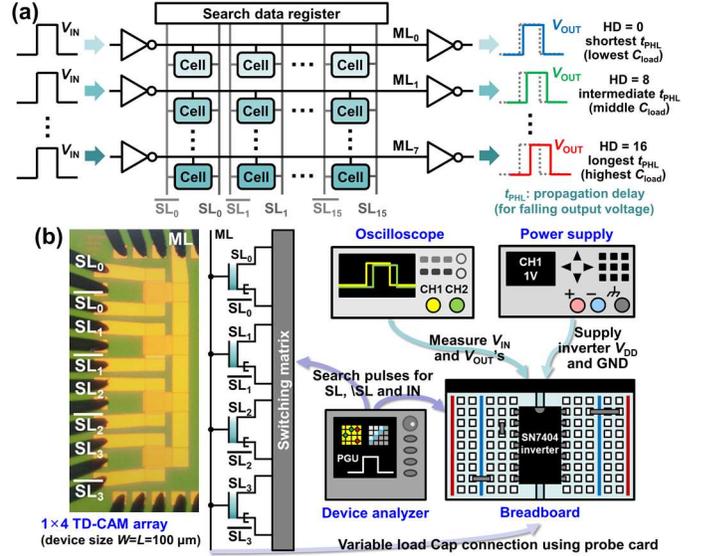

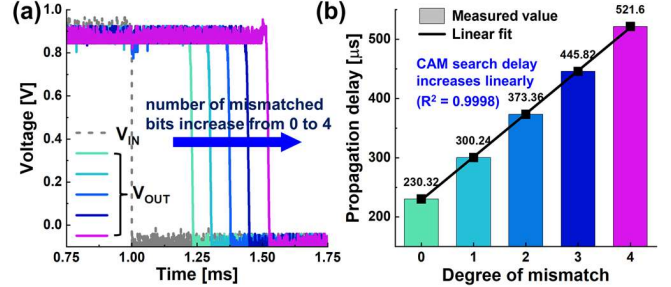

Fig. 3. (a) Design of the proposed TD CAM based on single ferroelectric memcapacitor. (b) Measurement environment of TD CAM search operation. The optical microscope image on the left is the top view of the fabricated 1×4 TD CAM array tipped with multi-pin probe card.

Fig. 4. (a) Measured transient waveforms of the proposed single TD CAM. (b) Measured TD CAM search delay which is linearly proportional to HD for the accurate evaluation of degree of mismatch.

Fig. 2c evaluates the variations in the measured capacitance and current [21]. The TD CAM cell exhibits less variation compared to the VD CAM cell, which leads to enhanced search reliability. The TFET *I-V* curves are sensitive to variations because their currents depend on the narrow tunneling region around the source and drain junctions. In contrast, the *C-V* properties are involved in the entire channel region where the inversion layer is formed [26], [27]. Table I summarizes the ML, SL, and \SL voltages ($V_{ML}$'s, $V_{SL}$'s, and $V_{\backslash SL}$'s) of the proposed TD CAM for the write and search operations. The cell capacitance states for bits "0" and "1" are updated by switching ferroelectric polarization via $V_{ML}$ pulses. In the search operation, depending on the inputs $V_{SL}$ and $V_{\backslash SL}$, the matched and mismatched cells add $C_{LCS}$ and $C_{HCS}$ to the ML load capacitance, respectively.

### B. TD CAM Measurement and Performance Prediction

To differentiate HDs for each ML based on the propagation delay, the proposed TD CAM incorporates single ferroelectric memcapacitor cells and two inverters per ML, as shown in Fig. 3a. The output of the first and the input of the second inverters are connected to the ML. The input of the first and the output of the second inverters are connected to the CAM input and output voltages ($V_{IN}$'s and $V_{OUT}$'s), respectively. In this

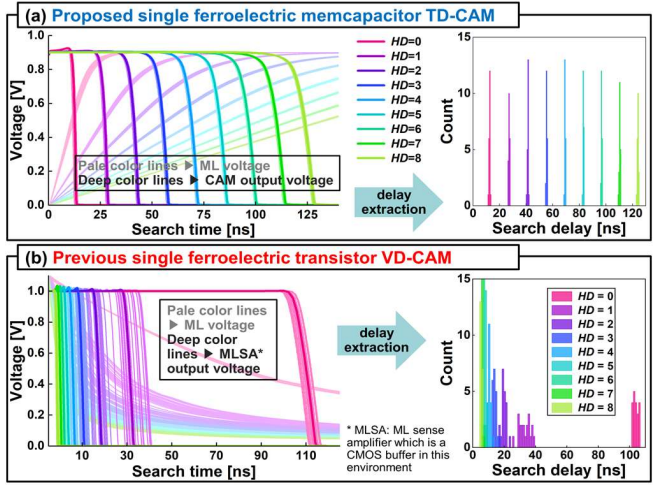

Fig. 5. Simulated transient response of $V_{ML}$'s and $V_{OUT}$'s during search operation (left) and extracted statistical distribution of search delay (right) evaluated for (a) proposed TD CAM and (b) previous VD CAM.

TABLE II
AREA COMPARISON OF THE PROPOSED AND THE PREVIOUS TD CAMs

| TD-CAM cell structure | 5T1C [b] [6], [7], [8] | 3T [9], [10] | 1C This work, [21] |
|---|---|---|---|
| Cell area [a] | 304 $F^2$ (~ 5.43 ×) | 200 $F^2$ (~ 3.57 ×) | 56 $F^2$ (1 ×) |

[a] F = half pitch. [b] T and C denote transistor and capacitor, respectively.

## III. SUMMARY

A novel, high-density, and accurate TD CAM based on a single ferroelectric memcapacitor with symmetric $C$-$V$ curves has been proposed. The proposed TD CAM demonstrates linear HD computation through modulation of the propagation delay. The proposed TD CAM outperforms previous single-transistor VD CAMs in terms of linearity and resilience to variation, achieving enhanced NN search accuracy and sensing margin. Moreover, the TD CAM offers superior energy efficiency and digital circuit compatibility. Furthermore, the proposed TD CAM achieves superior area efficiency compared to previous ferroelectric TD CAMs. Consequently, the proposed TD CAM is positioned as a promising solution for the performance and density limitations of the existing CAMs in on-device approximate search applications.

configuration, the tuned amount of ML load capacitance is directly proportional to the degree of mismatch. Consequently, the propagation delay from $V_{IN}$ to $V_{OUT}$ increases linearly in relation to the HD. Fig. 3b shows the measurement setup where the ML of the fabricated 1×4 TD CAM array is tied to inverters, and the Keysight B1500A semiconductor device analyzer and E5250A switching matrix supply signals.

Fig. 4 plots the measured search operation results of the proposed TD CAM. A linear correlation between the high-to-low propagation delay and the HD is experimentally verified. To evaluate search performance in larger arrays, Fig. 5 illustrates the SPICE Monte Carlo simulation results for the transient responses of 16-bit TD and VD CAM words [28]. The calibrated compact device model described in our previous works was applied by incorporating the variation parameters in Fig. 2c [21], [29]. The proposed TD CAM demonstrates higher reliability compared to the previous VD CAM for identifying HDs from 0 to 8 in 16-bit CAM words, assuming both utilize time-to-digital converters with equivalent resolution limits. The TD CAM achieves highly linear HD computing, offering narrower search delay distributions and improved sensing margins. In contrast, VD CAM suffers from nonlinear behavior and vulnerability to device variations, compromising the NN search accuracy.

Table II indicates that the proposed TD CAM achieves the highest integration density among existing ferroelectric TD CAMs [6]-[10]. Previous TD CAMs based on 5T1C and 3T architectures utilize ferroelectric memories as transistors rather than capacitors, requiring additional external capacitors or inverters per cell to separate each delay unit for linear TD operation. This requirement increases the cell size. In contrast, our proposed TD CAM employs ferroelectric memcapacitors as variable-capacitance load cells, optimizing area efficiency.